\documentclass[11pt]{article}
\usepackage{amsmath,amsfonts,feynmf}
\begin{document}

\date{\today}
\title{Runge-Kutta methods and renormalization}

\author{ Ch. Brouder\\
  \small Laboratoire de min\'eralogie cristallographie\\
  \small Universit\'es Paris 6, Paris 7, IPGP, Case 115, 4 place Jussieu\\
  \small 75252 Paris {\sc cedex} 05, France}
\maketitle
\abstract{A connection between the algebra of rooted trees used in renormalization
theory and Runge-Kutta methods is pointed out. Butcher's group and B-series are
shown to provide a suitable framework for renormalizing a toy model of field theory,
following Kreimer's approach. Finally B-series are used 
to solve a class of non-linear partial differential equations.}
\newcommand{\sixj}[6]{\left\{ \begin{array}{ccc}#1 &#2 &#3\\
                             #4 &#5 &#6\end{array} \right\} }
\newcommand{\threej}[6]{\left( \begin{array}{ccc}#1 &#2 &#3\\
                             #4 &#5 &#6\end{array} \right) }

\unitlength=1mm
\begin{fmffile}{essai1}


\newcommand{\tun}{\parbox{3mm}{
\begin{fmfchar*}(3,3)
  \fmfforce{(0.5w,0.5h)}{i1}
  \fmfv{decor.shape=circle,decor.filled=1,decor.size=30}{i1}
\end{fmfchar*}}}


\newcommand{\tdeux}{\parbox{3mm}{
\begin{fmfchar*}(3,5)
  \fmfforce{(0.5w,0.9h)}{i1}
  \fmfforce{(0.5w,0.1h)}{o1}
  \fmf{vanilla}{i1,o1}
  \fmfv{decor.shape=circle,decor.filled=1,decor.size=30}{i1}
  \fmfv{decor.shape=circle,decor.filled=0,decor.size=30}{o1}
\end{fmfchar*}}}


\newcommand{\ttroisun}{
\parbox{3mm}{
\begin{fmfchar*}(3,8)
  \fmfforce{(0.5w,0.9h)}{i1}
  \fmfforce{(0.5w,0.1h)}{o1}
  \fmf{vanilla}{i1,v1}
  \fmf{vanilla}{v1,o1}
  \fmfv{decor.shape=circle,decor.filled=1,decor.size=30}{i1}
  \fmfv{decor.shape=circle,decor.filled=0,decor.size=30}{v1}
  \fmfv{decor.shape=circle,decor.filled=0,decor.size=30}{o1}
\end{fmfchar*}
}}


\newcommand{\ttroisdeux}{
\parbox{5mm}{
\begin{fmfchar*}(5,6)
  \fmfforce{(0.5w,0.9h)}{i1}
  \fmfforce{(0.1w,0.1h)}{o1}
  \fmfforce{(0.9w,0.1h)}{o2}
  \fmf{vanilla}{i1,o1}
  \fmf{vanilla}{i1,o2}
  \fmfv{decor.shape=circle,decor.filled=1,decor.size=30}{i1}
  \fmfv{decor.shape=circle,decor.filled=0,decor.size=30}{o1}
  \fmfv{decor.shape=circle,decor.filled=0,decor.size=30}{o2}
\end{fmfchar*}
}}


\newcommand{\tquatreun}{
\parbox{3mm}{
\begin{fmfchar*}(3,12)
  \fmfforce{(0.5w,0.9h)}{i1}
  \fmfforce{(0.5w,0.1h)}{o1}
  \fmf{vanilla}{i1,v1}
  \fmf{vanilla}{v1,v2}
  \fmf{vanilla}{v2,o1}
  \fmfv{decor.shape=circle,decor.filled=1,decor.size=30}{i1}
  \fmfv{decor.shape=circle,decor.filled=0,decor.size=30}{v1}
  \fmfv{decor.shape=circle,decor.filled=0,decor.size=30}{v2}
  \fmfv{decor.shape=circle,decor.filled=0,decor.size=30}{o1}
\end{fmfchar*}
}}


\newcommand{\tquatredeux}{
\parbox{6mm}{
\begin{fmfchar*}(6,10)
  \fmfforce{(0.5w,0.9h)}{i1}
  \fmfforce{(0.1w,0.1h)}{o1}
  \fmfforce{(0.9w,0.1h)}{o2}
  \fmfforce{(0.5w,0.45h)}{v1}
  \fmf{vanilla}{i1,v1}
  \fmf{vanilla}{v1,o1}
  \fmf{vanilla}{v1,o2}
  \fmfv{decor.shape=circle,decor.filled=1,decor.size=30}{i1}
  \fmfv{decor.shape=circle,decor.filled=0,decor.size=30}{v1}
  \fmfv{decor.shape=circle,decor.filled=0,decor.size=30}{o1}
  \fmfv{decor.shape=circle,decor.filled=0,decor.size=30}{o2}
\end{fmfchar*}
}}


\newcommand{\tquatretrois}{
\parbox{6mm}{
\begin{fmfchar*}(6,10)
  \fmfforce{(0.5w,0.9h)}{i1}
  \fmfforce{(0.1w,0.5h)}{o1}
  \fmfforce{(0.1w,0.1h)}{o2}
  \fmfforce{(0.9w,0.5h)}{o3}
  \fmf{vanilla}{i1,o1}
  \fmf{vanilla}{o1,o2}
  \fmf{vanilla}{i1,o3}
  \fmfv{decor.shape=circle,decor.filled=1,decor.size=30}{i1}
  \fmfv{decor.shape=circle,decor.filled=0,decor.size=30}{o1}
  \fmfv{decor.shape=circle,decor.filled=0,decor.size=30}{o2}
  \fmfv{decor.shape=circle,decor.filled=0,decor.size=30}{o3}
\end{fmfchar*}
}}


\newcommand{\tquatrequatre}{
\parbox{8mm}{
\begin{fmfchar*}(8,6)
  \fmfforce{(0.5w,0.9h)}{i1}
  \fmfforce{(0.1w,0.1h)}{o1}
  \fmfforce{(0.5w,0.1h)}{o2}
  \fmfforce{(0.9w,0.1h)}{o3}
  \fmf{vanilla}{i1,o1}
  \fmf{vanilla}{i1,o2}
  \fmf{vanilla}{i1,o3}
  \fmfv{decor.shape=circle,decor.filled=1,decor.size=30}{i1}
  \fmfv{decor.shape=circle,decor.filled=0,decor.size=30}{o1}
  \fmfv{decor.shape=circle,decor.filled=0,decor.size=30}{o2}
  \fmfv{decor.shape=circle,decor.filled=0,decor.size=30}{o3}
\end{fmfchar*}
}}

\section{Introduction}

The purpose of this paper is to point out a link between two
apparently remote concepts: renormalization and Runge-Kutta
methods.

Renormalization enables us to remove infinities from quantum
field theory. Recently, Kreimer discovered a Hopf algebra
of rooted trees
that brings order and beauty in the intricate combinatorics of
renormalization \cite{Kreimer98}. He established formulas
that automate the subtraction of infinities to all orders
of the perturbation expansion, and proved the effectiveness
of his method for the practical computation of renormalized
quantities in joint works with Broadhurst \cite{Broadhurst}
and Delbourgo \cite{Delbourgo}. Moreover, his approach shines
new light on the problem of overlapping divergences
\cite{KreimerOD,Krajewski} and on the mechanics of the 
renormalization
group \cite{Kreimer}. Furthermore, Connes and Kreimer revealed
a deep connection between the algebra of rooted trees
(ART) and a Hopf algebra of diffeomorphisms \cite{Connes}.

On the other hand, Runge published in 1895 \cite{Runge}
an efficient algorithm to compute the solution of 
ordinary differential equations. For an equation of
the type $dy/ds=f(y(s))$, he defines
recursively $k_1=f(y_n)$, $k_2=f(y_n+h k_1/2)$,
$y_{n+1}=y_n+hk_2$. His algorithm was improved
in 1901 by Kutta, and became known as the Runge-Kutta
method. It is now one of the most widely used
numerical methods.

In 1972, Butcher published an extraordinary article
where he analyzed general Runge-Kutta methods on the basis
of the ART. He showed that the Runge-Kutta methods form
a group\footnote{Hairer and Wanner called it the
Butcher group \cite{Hairer74}.} and found  
explicit expressions for the inverse
of a method or the product of two methods. He also
defined sums over trees that are now called B-series
in honour of Butcher.

Altough the Hopf algebra structure of ART is implicit
all along his paper, Butcher did not mention
it\footnote{The opposite of the present Hopf structure of ART 
was discussed in 1986 by 
D\"ur (\cite{Dur}, p.88-90), together with
the corresponding Lie algebra (identical, up to
a sign, with the one defined in \cite{Connes}).}.
Important developments were made in 1974 by Hairer
and Wanner \cite{Hairer74}. Since then, B-series
are used routinely in the analysis of Runge-Kutta
methods.

Our main purpose is to show that the results and concepts
established by Kreimer fit nicely into the Runge-Kutta language,
and that the tools developed by Butcher have a range of application
much wider than the numerical analysis of ordinary differential
equations.

The present expository paper will be reasonably self-contained.
After an introduction to rooted trees, the genetic relation
between ART and differentials is presented. Then Butcher's 
approach to Runge-Kutta methods is sketched. Several B-series
are calculated and the connection with the Hopf structure
of ART is exhibited. The application of Runge-Kutta methods
to renormalization is exposed using a toy model which is
solved non perturbatively. Finally, the solution of non-linear
partial differential equations is written as a formal
B-series.

\section{The rooted trees}

A rooted tree is a graph with a designated vertex called
a root such that there is a unique path from the root
to any other vertex in the tree \cite{Tucker}. Several examples of
rooted trees are given in the appendix, where the root
is the black point and the other vertices are white points
(the root is at the top of the tree).
The length of the unique path from a vertex $v$ to the root 
is called the level number of vertex $v$. The root has level
number 0. For any vertex $v$
(except the root), the father of $v$ is the unique vertex $v'$
with an edge common with $v$ and a smaller level number.
Conversely, $v$ is a son of $v'$. A vertex with no sons is
a leaf.
Rooted trees are sometimes called pointed trees or arborescences.

The tree with one vertex is $\tun$, the ``tree'' with
zero vertex is designated by $1$.

\subsection{Operations and functions on trees}

An important operation is the merging of trees.
If $t_1$,\dots,$t_k$ are trees, $t=B^+(t_1,t_2,\dots,t_k)$
is defined as the tree obtained by creating a new vertex $r$
and by joining the roots of $t_1$,\dots,$t_k$ to $r$, which
becomes the root of $t$. This operation is also denoted
by $t=[t_1,t_2,\dots,t_k]$, but we avoid this notation because
of the possible confusion with commutators.

In \cite{Connes}, Connes and Kreimer defined a natural growth
operator $N$ on trees: $N(t)$ is the set of $|t|$ trees
$t_i$, where each $t_i$ is a tree with $|t|+1$ vertices
obtained by attaching an additional leaf to a vertex of 
$t$. For example $N(1)=\tun$, 
\begin{eqnarray*}
N(\tun)&=&\tdeux,\quad N(\tdeux)=\ttroisun+\ttroisdeux\quad
N(\ttroisdeux)=\tquatrequatre+2\,\tquatretrois.
\end{eqnarray*}
Some trees may appear with multiplicity.

A number of functions on rooted trees have been defined 
independently by several authors:

$|t|$ designates the number of vertices of a tree $t$
(alternative notation is $r(t)$, $\rho(t)$ and $\# t$).
Clearly, 
$|B^+(t_1,t_2,\dots,t_k)|=|t_1|+|t_2|+\cdots+|t_k|+1$.

The tree factorial $t!$ is defined recursively as
\begin{eqnarray*}
\tun !&=&1\\
B^+(t_1,t_2,\dots,t_k)!&=&|B^+(t_1,t_2,\dots,t_k)|\, t_1! t_2! \cdots  t_k!.
\end{eqnarray*}
(an alternative notation is $\gamma(t)$).
The notation $t!$ is taken from
Kreimer \cite{Kreimer} because $t!$ generalizes the factorial of a number.
Besides $t!$ has also similarities 
with the product of hooklengthes of a Young diagram in the representation theory
of the symmetric group \cite{Fomin}. 
A few examples may be useful
\begin{eqnarray*}
\tdeux !&=&2,\quad \ttroisun !=6,\quad \tquatredeux != 12,\quad  \tquatreun != 24,
\quad\tquatrequatre !=4.
\end{eqnarray*}

\subsection{On $CM(t)$}
$CM(t)$ was defined in \cite{Kreimer} as the number of
times tree $t$ appears in $N^n(1)$ where $n=|t|$ is the number
of vertices of $t$. In the literature (\cite{Butcher63},
\cite{Butcher}, p.92, \cite{Hairer}, p.147), 
$CM(t)$ is written as
$\alpha(t)$ and considered as the number of ``heap-ordered trees''
with shape $t$, where a heap-ordered tree with shape $t$ is
a labelling of each vertex of $t$ (i.e. a bijection between
the vertices and the set of numbers $0,1,\dots |t|-1$) such
that the labels decrease along the path going from any vertex
to the root. This is called a monotonic labelling in
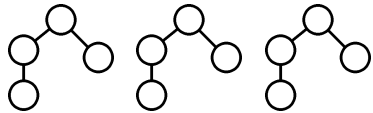
\cite{Hairer}, p.147.

\begin{eqnarray*}
\parbox{10mm}{
\begin{fmfchar*}(10,10)
  \fmftop{i1}
  \fmfforce{(0,0)}{o2}
  \fmfforce{(0,0.6h)}{o1}
  \fmfright{o3}
  \fmf{vanilla}{i1,o1}
  \fmf{vanilla}{o1,o2}
  \fmf{vanilla}{i1,o3}
  \fmfv{decor.shape=circle,decor.filled=0,%
   label=\mbox{\small$0$},label.angle=0,label.dist=0,decor.size=90}{i1}
  \fmfv{decor.shape=circle,decor.filled=0,%
   label=\mbox{\small$1$},label.angle=0,label.dist=0,decor.size=90}{o1}
  \fmfv{decor.shape=circle,decor.filled=0,%
   label=\mbox{\small$2$},label.angle=0,label.dist=0,decor.size=90}{o2}
  \fmfv{decor.shape=circle,decor.filled=0,%
   label=\mbox{\small$3$},label.angle=0,label.dist=0,decor.size=90}{o3}
\end{fmfchar*}
}
\quad\quad
\parbox{10mm}{
\begin{fmfchar*}(10,10)
  \fmftop{i1}
  \fmfforce{(0,0)}{o2}
  \fmfforce{(0,0.6h)}{o1}
  \fmfright{o3}
  \fmf{vanilla}{i1,o1}
  \fmf{vanilla}{o1,o2}
  \fmf{vanilla}{i1,o3}
  \fmfv{decor.shape=circle,decor.filled=0,%
   label=\mbox{\small$0$},label.angle=0,label.dist=0,decor.size=90}{i1}
  \fmfv{decor.shape=circle,decor.filled=0,%
   label=\mbox{\small$2$},label.angle=0,label.dist=0,decor.size=90}{o1}
  \fmfv{decor.shape=circle,decor.filled=0,%
   label=\mbox{\small$3$},label.angle=0,label.dist=0,decor.size=90}{o2}
  \fmfv{decor.shape=circle,decor.filled=0,%
   label=\mbox{\small$1$},label.angle=0,label.dist=0,decor.size=90}{o3}
\end{fmfchar*}
}
\quad\quad
\parbox{10mm}{
\begin{fmfchar*}(10,10)
  \fmftop{i1}
  \fmfforce{(0,0)}{o2}
  \fmfforce{(0,0.6h)}{o1}
  \fmfright{o3}
  \fmf{vanilla}{i1,o1}
  \fmf{vanilla}{o1,o2}
  \fmf{vanilla}{i1,o3}
  \fmfv{decor.shape=circle,decor.filled=0,%
   label=\mbox{\small$0$},label.angle=0,label.dist=0,decor.size=90}{i1}
  \fmfv{decor.shape=circle,decor.filled=0,%
   label=\mbox{\small$1$},label.angle=0,label.dist=0,decor.size=90}{o1}
  \fmfv{decor.shape=circle,decor.filled=0,%
   label=\mbox{\small$3$},label.angle=0,label.dist=0,decor.size=90}{o2}
  \fmfv{decor.shape=circle,decor.filled=0,%
   label=\mbox{\small$2$},label.angle=0,label.dist=0,decor.size=90}{o3}
\end{fmfchar*}
}
\end{eqnarray*}

There are $(n-1)!$ heap-ordered trees with $n$ vertices.
This can be seen by a recursive argument. Take a heap-ordered
tree $t$ with $n$ vertices, and make $n$ labeled trees by
adding a new vertex with label $n$ to each vertex of $t$.
Then all the created trees are heap-ordered (because the 
added vertex is a leaf and all other labels are smaller
than $n$). Furthermore, all the heap-ordered trees
created by this process from the set of heap-ordered trees with $n$
vertices are different. Therefore, there are at least
$n!$ heap-ordered trees with $n+1$ vertices. 
On the other hand, in each heap-ordered tree $t$
with $n+1$ vertices, the vertex labeled $n$ is a leaf,
therefore $t$ can be created from a heap-ordered tree with
$n$ vertices by adding this leaf with label $n$.
So there are exactly $n!$ heap-ordered trees with $n+1$ vertices.
We shall give a non combinatiorial proof of this fact in the sequel.

Since $N(t)$ is defined by the addition
of a leaf to all the vertices of $t$,
$\alpha(t)$  is the number of heap-ordered trees with shape $t$.
This number has been calculated in \cite{Butcher63}
(see also. \cite{Butcher}, p.92):
\begin{eqnarray*}
\alpha(t)=\frac{|t|!}{t!S_t},
\end{eqnarray*}
where $S_t$ is the symmetry factor of $t$, defined in
\cite{Broadhurst,Kreimer} and in \cite{Butcher}
where it is denoted by $\sigma(t)$.

Note that there is a simple correspondence between
the permutations of $n-1$ numbers and the heap-ordered
trees. Let $(p_1,\dots,p_{n-1})$ be a permutation
of $(1,\dots,n-1)$, then
\begin{itemize}
\item $p_1$ is a subroot, labeled $p_1$
\item for $i$=2 to $n-1$
\begin{itemize}
\item if all $p_j$ for $1\le j\le i$ are such that
  $p_j > p_i$, then $p_i$ is a subroot, labeled $p_i$
\item otherwise, let $p_j$ be first number such that 
  $p_j < p_i$, in the series
  $p_{i-1},p_{i-2},\dots,p_1$, then the $i$-th vertex,
  labeled $p_i$, is linked to $p_j$ by a line
\end{itemize}
\item when all $(p_1,\dots,p_{n-1})$ have been processed,
  all subroots are linked to a common root, labeled $0$
\end{itemize}

On the other hand, starting from a heap-ordered tree $t$,
$t$ is arranged so that the set of all vertices
with a given level number are ordered with
labels increasing from right to left. Then the permutation
is built by gathering the labels  through  a depth-first search 
(backtracking) of the tree from left to right.
For instance, the permutation corresponding to the
three labeled trees of the above example are
(312), (231) and (213).

Finally, we use the term algebra of rooted trees and not
Hopf algebra of rooted trees because, thanks to the work
of Butcher, the Hopf structure is only one aspect of the
ART.

\section{Differentials and rooted trees}


Assume that we want to solve the equation 
$(d/ds)x(s)=F[x(s)]$, $x(s_0)=x_0$,
where $s$ is a real, $x$ is in $\mathbb{R}^N$ 
and $F$ is a smooth
function from $\mathbb{R}^n$ to $\mathbb{R}^N$,
with components $f^i(x)$. This is the equation
of the flow of a vector field.

\subsection{Calculation of the $n$-th derivative}
Let us write the derivatives of the $i$-th component of
$x(s)$ with respect to $s$:

\begin{eqnarray*} 
\frac{d^2x^i(s)}{ds^2}&=&\frac{d}{ds} f^i[x(s)] 
= \sum_j\frac{\partial f^i}{\partial x_j}[x(s)] \frac{dx^j}{ds} 
= \sum_j\frac{\partial f^i}{\partial x_j}[x(s)] f^j[x(s)]
\end{eqnarray*} 

\begin{eqnarray*} 
\frac{d^3x^i(s)}{ds^3}&=&\frac{d}{ds} 
  \left(\sum_j\frac{\partial f^i}{\partial x_j}[x(s)] f^j[x(s)]\right)\\
&=& \sum_{jk}\frac{\partial^2 f^i}{\partial x_j\partial x_k}[x(s)]
   f^j[x(s)] f^k[x(s)] +
 \sum_{jk}\frac{\partial f^i}{\partial x_j}[x(s)] 
          \frac{\partial f^j}{\partial x_k}[x(s)] f^k[x(s)].
\end{eqnarray*} 

A simplified notation is now required.  Let 
\begin{eqnarray*}
f^i &=& f^i[x(s)]  \\
f^i_{j_1j_2\cdots j_k} &=&
\frac{\partial^k f^i}{\partial x_{j_1}\cdots\partial x_{j_k}}[x(s)],
\end{eqnarray*}
so that
\begin{eqnarray*} 
\frac{dx^i(s)}{ds}&=& f^i\quad
\frac{d^2x^i(s)}{ds^2}= f^i_j f^j\quad
\frac{d^3x^i(s)}{ds^3}= f^i_{jk} f^j f^k + f^i_j f^j_k f^k,
\end{eqnarray*} 
where summation over indices appearing in lower and upper
positions is implicitly assumed.

With this notation, we can write the next term as
\begin{eqnarray*}
\frac{d^4x^i(s)}{ds^4}&=& f^i_j f^j_k f^k_l f^l 
 + f^i_j f^j_{kl} f^k f^l + 3 f^i_{jk} f^j_l f^k f^l
 + f^i_{jkl} f^j f^k f^l \\
&=&
\parbox{10mm}{
\begin{fmfchar*}(10,12)
  \fmftop{i1}
  \fmfbottom{o1}
  \fmf{vanilla}{i1,v1}
  \fmf{vanilla}{v1,v2}
  \fmf{vanilla}{v2,o1}
  \fmfv{decor.shape=circle,decor.filled=1,%
   label=\mbox{\small$i$},label.angle=0,label.dist=20,decor.size=30}{i1}
  \fmfv{decor.shape=circle,decor.filled=0,%
   label=\mbox{\small$j$},label.angle=0,label.dist=20,decor.size=30}{v1}
  \fmfv{decor.shape=circle,decor.filled=0,%
   label=\mbox{\small$k$},label.angle=0,label.dist=20,decor.size=30}{v2}
  \fmfv{decor.shape=circle,decor.filled=0,%
   label=\mbox{\small$l$},label.angle=0,label.dist=20,decor.size=30}{o1}
\end{fmfchar*}
}
\quad
\quad
\quad
\parbox{6mm}{
\begin{fmfchar*}(6,10)
  \fmftop{i1}
  \fmfbottom{o1,o2}
  \fmfforce{(0.5w,0.5h)}{v1}
  \fmf{vanilla}{i1,v1}
  \fmf{vanilla}{v1,o1}
  \fmf{vanilla}{v1,o2}
  \fmfv{decor.shape=circle,decor.filled=1,%
   label=\mbox{\small$i$},label.angle=0,label.dist=20,decor.size=30}{i1}
  \fmfv{decor.shape=circle,decor.filled=0,%
    label=\mbox{\small$j$},decor.size=30}{v1}
  \fmfv{decor.shape=circle,decor.filled=0,%
    label=\mbox{\small$k$},decor.size=30}{o1}
  \fmfv{decor.shape=circle,decor.filled=0,%
    label=\mbox{\small$l$},decor.size=30}{o2}
\end{fmfchar*}
}
\quad
\quad
\quad
\quad
\parbox{6mm}{
\begin{fmfchar*}(6,10)
  \fmftop{i1}
  \fmfforce{(0,0)}{o2}
  \fmfforce{(0,0.5h)}{o1}
  \fmfright{o3}
  \fmf{vanilla}{i1,o1}
  \fmf{vanilla}{o1,o2}
  \fmf{vanilla}{i1,o3}
  \fmfv{decor.shape=circle,decor.filled=1,%
   label=\mbox{\small$i$},label.angle=0,label.dist=20,decor.size=30}{i1}
  \fmfv{decor.shape=circle,decor.filled=0,%
   label=\mbox{\small$j$},decor.size=30}{o1}
  \fmfv{decor.shape=circle,decor.filled=0,%
   label=\mbox{\small$l$},decor.size=30}{o2}
  \fmfv{decor.shape=circle,decor.filled=0,%
   label=\mbox{\small$k$},decor.size=30}{o3}
\end{fmfchar*}
}
\quad
\quad
\quad
\quad
\parbox{8mm}{
\begin{fmfchar*}(8,6)
  \fmftop{i1}
  \fmfstraight
  \fmfbottom{o1,o2,o3}
  \fmf{vanilla}{i1,o1}
  \fmf{vanilla}{i1,o2}
  \fmf{vanilla}{i1,o3}
  \fmfv{decor.shape=circle,decor.filled=1,%
   label=\mbox{\small$i$},decor.size=30}{i1}
  \fmfv{decor.shape=circle,decor.filled=0,%
   label=\mbox{\small$j$},decor.size=30}{o1}
  \fmfv{decor.shape=circle,decor.filled=0,%
   label=\mbox{\small$k$},decor.size=30}{o2}
  \fmfv{decor.shape=circle,decor.filled=0,%
   label=\mbox{\small$l$},decor.size=30}{o3}
\end{fmfchar*}
}
\end{eqnarray*}

This relation between differentials and rooted tree was
established by Arthur Cayley in 1857 \cite{Cayley}.
With this notation, there is a one-to-one relation
between a rooted tree with $n$ vertices and a 
term of $d^n x(s)/ds^n$

\subsection{Elementary differentials}
A little bit more formally, we can follow Butcher
(\cite{Butcher}, p.154.) and call
``elementary differentials'' the 
$\delta_t$ defined recursively for each rooted tree $t$ by:
\begin{eqnarray}
\delta^i_\bullet &=& f^i \nonumber\\
\delta^i_t &=& 
f^i_{j_1j_2\cdots j_k}  \delta^{j_1}_{t_1}
\delta^{j_2}_{t_2} \cdots \delta^{j_k}_{t_k}
\quad\mathrm{when}\quad t=B^+(t_1,t_2,\cdots,t_k). \label{defdeltat}
\end{eqnarray}

Using this correspondence between rooted trees and
differential expressions, we establish the identity:
\begin{eqnarray*}
N\delta_{t}\equiv\frac{d\delta_t}{ds}=\delta_{N(t)},
\end{eqnarray*}
where $N(t)$ is the natural growth operator of rooted
trees defined in Ref.\cite{Connes}.

So that the solution of the flow equation is
\begin{eqnarray}
x(s)&=&x_0+\int_{s_0}^s ds' \exp[s'N]\delta_\bullet\nonumber\\
&=&x_0+\sum_t \frac{(s-s_0)^{|t|}}{|t|!}
  \alpha(t) \delta_t(s_0), \label{Butcherflow}
\end{eqnarray}
where $|t|$ is the number of vertices of $t$,
and $\alpha(t)$ is called $CM(t)$ in \cite{Kreimer}.

\section{Runge-Kutta methods}
We shall see that sum over trees appear quite naturally
with differential equations. So, if one is given a
function $\phi$ that assigns a value (e.g. a real,
a complex, a vector) to each tree $t$, is there
a function $f$ such that $\phi(t)=\delta_t$.
Generally, the answer is no. Consider a function $\phi$ such that
all components are equal (and denoted also by $\phi$):
\begin{eqnarray*}
\phi(\tun)&=&1,\quad \phi(\tdeux)=a,\quad \phi(\ttroisun)=b, 
\end{eqnarray*}
so that for any $i$, $f^i=1$, $f^i_j f^j=a$ and $f^i_j f^j_kf^k=b$.
The first two equations give $\sum_jf^i_j=a$, so that
the third gives 
$f^i_j f^j_kf^k=\sum_j f^i_j a=a^2$, and $\phi$ cannot
be represented as elementary differentials 
(i.e. it cannot be the $\delta_t$) of a function $f$ if $b\not=a^2$.
In fact, the number of functions reachable as
elementary differentials is rather narrow.

Given such a function $\phi$ over rooted trees, we extend it
to a homomorphism of the algebra of rooted trees by linearity
and 
$\phi(tt')=\phi(t)\phi(t')$ where the componentwise product was
used on the right-hand side.
If vector flows are not enough to span all possible $\phi$,
what more general equation can do that? As we shall see now,
the answer is the Runge-Kutta methods.

\subsection{Butcher's approach to the Runge-Kutta methods}
To solve a flow equation 
$dx(s)/ds=F[x(s)]$, 
some efficient numerical algorithms are known as
Runge-Kutta methods. They are determined by
a $m\times m$ matrix $a$ and an $m$-dimensional vector $b$, and
at each step a vector $x_n$ is defined as a function
of the previous value $x_{n-1}$ by:
\begin{eqnarray*}
X_i &=& x_{n-1}+h\sum_{j=1}^m a_{ij} F(X_j)\\
x_n &=& x_{n-1}+h\sum_{j=1}^m b_j F(X_j),
\end{eqnarray*}
where $i$ range from $1$ to $m$.
If the matrix $a$ is such that $a_{ij}=0$ if
$j\ge i$ then the method is called explicit (because
each $X_i$ can be calculated explicitly), otherwise
the method is implicit. 

In 1963, Butcher showed that the solution of the
corresponding equations:
\begin{eqnarray*}
X_i(s) &=& x_0+(s-s_0)\sum_{j=1}^m a_{ij} F(X_j(s))\\
x(s) &=& x_0+(s-s_0)\sum_{j=1}^m b_j F(X_j(s)),
\end{eqnarray*}
is given by
\begin{eqnarray}
X_i(s)&=&x_0+\sum_t \frac{(s-s_0)^{|t|}}{|t|!}
  \alpha(t) t! \sum_{j=1}^m a_{ij} \phi_j(t) \delta_t(s_0) \nonumber\\
x(s)&=&x_0+\sum_t \frac{(s-s_0)^{|t|}}{|t|!}
  \alpha(t) t! \phi(t) \delta_t(s_0). \label{Bseries}
\end{eqnarray}

These series over trees are called B-series in the numerical
analysis literature, in honour of John Butcher (\cite{Hairer}, p.264).
The homomorphism $\phi$ is defined recursively as a function
of $a$ and $b$, for $i=1,\dots,m$:
\begin{eqnarray*}
\phi_i(\tun)&=&1\\
\phi_i(B^+(t_1\cdots t_k))&=&\sum_{j_1,\dots,j_k}
  a_{ij_1}\dots a_{ij_k}\phi_{j_1}(t_1)\dots\phi_{j_k}(t_k)\\
\phi(t)&=&\sum_{i=1}^m b_i \phi_i(t).
\end{eqnarray*}
Comparing Eqs.(\ref{Butcherflow}) and (\ref{Bseries})
it is clear that the Runge-Kutta approximates the solution
of the original flow equation up to order $n$ 
if $\phi(t)=1/t!$ for all trees with up to $n$ vertices.

In 1972 \cite{Butcher72}, Butcher made further progress.
Firstly he showed that Runge-Kutta methods are ``dense'' in
the space of rooted tree homomorphisms. More precisely, he
showed that given any finite set of trees $T_0$ and any
function $\theta$ from $T_0$ to $\mathbb{R}$, then there is a
Runge-Kutta method (i.e. a matrix $a$ and a vector $b$) such
that the corresponding $\phi$ agrees with $\theta$ on
$T_0$ (see also \cite{Butcher} p.167).

\subsection{Further developments}
Furthermore, Butcher proved that the combinatorics he
used to study Runge-Kutta methods in 1963 \cite{Butcher63}
was hiding an algebra. If ($a$,$b$) and ($a'$,$b'$) are
two Runge-Kutta methods, with the corresponding 
homomorphisms $\phi$ and $\phi'$, then the product homomorphism is
defined (in Hopf algebra terms) by 
\begin{eqnarray*}
(\phi\star\phi')(t)&=&m[(\phi\otimes\phi')\Delta(t)].
\end{eqnarray*}

Butcher proved
that the $\phi$ derived from Runge-Kutta methods form
a group. Again, this is nicely interpreted within the
Hopf structure of the ART.
For instance,
the inverse of the element $\phi$ is simply defined
by $\phi^{-1}(t)=\phi[S(t)]$, where $S$ is the antipode.
This concept of inverse is quite important in practice since it
is involved in the concept of self-adjoint Runge-Kutta
methods, which have long-term stability in time-reversal symmetric
problems (\cite{Hairer}, p.219). 
The adjoint is defined within our approach by
$\phi^*(t)=(-1)^{|t|}\phi[S(t)]$.

On the other hand,
Butcher found an explicit expression for all the Hopf operations
of the ART. Given the method ($a$,$b$) for $\phi$, he expressed
the method ($a'$,$b'$) for $\phi\circ S$ ($S$ is the antipode) 
in (simple) terms of ($a$,$b$).
Moreover,
(\cite{Butcher}, p.312 et sq.), if ($a$,$b$) 
and ($a'$,$b'$) are
two Runge-Kutta methods (with dimensions $m$ and $m'$, respectively),
corresponding to $\phi$ and $\phi'$, the method  ($a"$,$b"$)
corresponding to the convolution product $(\phi\star\phi')$ is
\begin{eqnarray*}
a''_{ij}&=&a_{ij}\mathrm{\quad if\quad} 1\le i\le m
               \mathrm{\quad and\quad} 1\le j\le m,\\
a''_{ij}&=&a'_{ij}\mathrm{\quad if\quad} m+1\le i\le m+m'
               \mathrm{\quad and\quad} m+1\le j\le m+m',\\
a''_{ij}&=&b_{j}\mathrm{\quad if\quad} m+1\le i\le m+m'
               \mathrm{\quad and\quad} 1\le j\le m,\\
a''_{ij}&=&0 \mathrm{\quad if\quad} 1\le i\le m
               \mathrm{\quad and\quad} m+1\le j\le m+m',\\
b''_i &=& b_i \mathrm{\quad if\quad} 1\le i\le m,\\
b''_i &=& b'_i \mathrm{\quad if\quad} m+1\le i\le m+m'.
\end{eqnarray*}

In 1974, Hairer and Wanner (\cite{Hairer}, p.267) built upon the
work of Butcher and proved the following important result:
if we denote
\begin{eqnarray}
B(\phi,F)&=&1+\sum_t \frac{(s-s_0)^{|t|}}{|t|!}
  \alpha(t) t! \phi(t) \delta_t(s_0)  \label{Hairer1}
\end{eqnarray}
then
\begin{eqnarray}
B(\phi',B(\phi,F))&=&B(\phi\star\phi',F),  \label{Hairer2}
\end{eqnarray}
where $B(\phi',B(\phi,F))$ is the same as Eq.(\ref{Hairer1}),
with $\phi(t)$ replaced by $\phi'(t)$ and 
$\delta_t$ replaced by $\delta'(t)$
(i.e.  $\delta'(t)$ is calculate as $\delta_t$, but
 with the function $B(\phi,F)(s)$ instead of the function
  $F(x(s))$).

In other words, the group of homomorphisms acts on the right
on the functions $F$.

\section{The continuous limit}

In his seminal article \cite{Butcher72}, Butcher did not
restrict his treatment to finite sets of indices.
It is possible to consider the continuous limit of Runge-Kutta
methods. A possible form of it is an integral equation, which
we write artitrarily between 0 and 1:
\begin{eqnarray*}
X_u(s) &=& x_0+(s-s_0)\int_0^1 dv a(u,v) F(X_v(s))\\
x(s) &=& x_0+(s-s_0)\int_0^1 b(u) du F(X_u(s)),
\end{eqnarray*}
the solution of which are
\begin{eqnarray*}
X_u(s)&=&x_0+\sum_t \frac{(s-s_0)^{|t|}}{|t|!}
  \alpha(t) t! \int_0^1 dv a(u,v) \phi_v(t) \delta_t(s_0)\\
x(s)&=&x_0+\sum_t \frac{(s-s_0)^{|t|}}{|t|!}
  \alpha(t) t! \phi(t) \delta_t(s_0).
\end{eqnarray*}
The homomorphism $\phi$ is defined recursively as a function
of $a$ and $b$:
\begin{eqnarray*}
\phi_u(\tun)&=&1\\
\phi_u(B^+(t_1\cdots t_k))&=&\int_0^1 du_1 a(u,u_1) \phi_{u_1}(t_1)
       \dots\int_0^1 du_k a(u,u_k) \phi_{u_k}(t_k)\\
\phi(t)&=&\int_0^1 du b(u) \phi_u(t).
\end{eqnarray*}
Continuous RK-methods do not seem to have been much used, except for
an example in Butcher's book (\cite{Butcher} p.325).

\subsection{Butcher's example \label{Butcherexample}}
It will be useful in the following to have the results of
a modified version of Butcher's example.
So, we consider:
\begin{eqnarray}
X_u(s)&=&x_0+(s-s_0)\int_0^u F[X_v(s)] dv \label{Picard}\\
x(s)&=&x_0+(s-s_0)\int_0^1 F[X_u(s)] du\nonumber,
\end{eqnarray}
which corresponds to $a(u,v)=1_{[0,u]}(v)$, $b(u)=1$.
This Runge-Kutta method will be used again in the sequel, and
will be referred to as the ``simple integral method''.

If we take the derivative of Eq.(\ref{Picard}) with respect
to $u$ we obtain
\begin{eqnarray*}
\frac{d}{du} X_u(s)=(s-s_0) F[X_u(s)],
\end{eqnarray*}
so $X_u(s)=y(s_0+(s-s_0)u)$, where $y(s)$ is the solution
\begin{eqnarray*}
y(s)=x_0+\int_{s_0}^s F[y(s')] ds'.
\end{eqnarray*}
Moreover
\begin{eqnarray*}
x(s)&=&x_0+(s-s_0)\int_0^1 F[X_u(s)] du\\
     &=&x_0+(s-s_0)\int_0^1 F[y(s_0+(s-s_0)u)] du\\
     &=&x_0+\int_{s_0}^s F[y(s')] ds'  = y(s).
\end{eqnarray*}

The corresponding homomorphism $\phi(t)$ is defined by
\begin{eqnarray*}
\phi_u(\tun)&=&1\\
\phi_u(B^+(t_1\cdots t_k))&=&\int_0^u du_1 \phi_{u_1}(t_1)
       \dots\int_0^u du_k \phi_{u_k}(t_k)\\
\phi(t)&=&\int_0^1 du \phi_u(t).
\end{eqnarray*}
Using the facts that
$|B^+(t_1\cdots t_k)|=(|t_1|+\cdots+|t_k|+1)$ and
$B^+(t_1\cdots t_k)!=(|t_1|+\cdots+|t_k|+1)t_1!\dots t_k!$
it is proved that the solutions of these equations are
\begin{eqnarray*}
\phi_u(t)&=&\frac{|t|u^{|t|-1}}{t!}\\
\phi(t)&=&\frac{1}{t!}.
\end{eqnarray*}

If we introduce $\phi(t)=1/t!$ into Eq.(\ref{Bseries})
we obtain Eq.(\ref{Butcherflow}). So we confirm that
the solution of the equation
\begin{eqnarray*}
x(s)=x_0+\int_{s_0}^s F[x(s')] ds'
\end{eqnarray*}
is
\begin{eqnarray*}
x(s)&=&x_0+
\sum_t \frac{(s-s_0)^{|t|}}{|t|!}
  \alpha(t) \delta_t(s_0). 
\end{eqnarray*}

\subsection{First applications}

The above example can already bring some interesting
applications. But we must start by giving a way to
calculate $\delta_t(s_0)$ in a simple case.

\subsubsection{Calculation of  $\delta_t(s_0)$ \label{sectiondeltat}}

To obtain specific results, we must choose a particular
function $F$.  The simplest choice is to take a vector function
$F$, with components 
$f^i(x)=f(\sum_j x_j/N)$, where $N$ is the dimension of the
vector space and $f$ has the series expansion
\begin{eqnarray*}
f(s)=\sum_{n=0}^\infty \frac{f^{(n)}(0) s^n}{n!}.
\end{eqnarray*}

From the definition of $\delta_t$ in Eq.(\ref{defdeltat}),
one can show recursively that, for $i=1,\dots,N$,
$\delta^i_t(0)$ is independent of $i$ (and will be denoted
$\delta_t$) and 
\begin{eqnarray}
\delta_\bullet &=& f(0) \nonumber\\
\delta_t &=& f^{(k)}(0) \delta_{t_1} \delta_{t_2} \cdots \delta_{t_k}
\quad\mathrm{when}\quad t=B^+(t_1,t_2,\cdots,t_k). \label{defdeltatsimple}
\end{eqnarray}
In ref.\cite{Kreimer}, Kreimer defined a similar quantity,
that he called $B_t$. Here $\delta_t$ and $B_t$ will be
used as synonymous.

The simplest case is $f(s)=\exp s$ and $s_0=0$, where $f^{(n)}(0)=1$
and $\delta_t=1$ for all trees $t$.

\subsubsection{Weighted sum of rooted trees}

If we take $f=\exp$, $s_0=0$ and $x_0=0$ in Butcher's example
(see section \ref{Butcherexample}), we have to solve the equation
\begin{eqnarray*}
x(s)=\int_{0}^s \exp[x(s')] ds'
\end{eqnarray*}
which can be differentiated to give $x'(s)=\exp(x(s))$
with $x(0)=0$. This has the solution
\begin{eqnarray*}
x(s)=-\log(1-s)=\sum_{n=1}^\infty \frac{s^n}{n}.
\end{eqnarray*}
On the other hand, the corresponding homomorphism is
$\phi(t)=1/t!$ and the B-series for this problem is
\begin{eqnarray*}
x(s)&=&
\sum_t \frac{s^{|t|}}{|t|!}
  \alpha(t).
\end{eqnarray*}
Comparing the last two results, we find
\begin{eqnarray*}
\sum_{|t|=n} \alpha(t)=(n-1)!
\end{eqnarray*}
in other words, the number of heap-ordered trees
with $n$ vertices is $(n-1)!$.

\subsubsection{Derivative of inverse functions}

We can try to extend the last example to an arbitrary
function $f(x)$.
The equation to solve becomes
\begin{eqnarray*}
x(s)=\int_{0}^s f[x(s')] ds',
\end{eqnarray*}
or $x'(s)=f(x(s))$ with $x(0)=0$.
Let 
\begin{eqnarray*}
S(x)=\int_{0}^x \frac{dy}{f(y)},
\end{eqnarray*}
which gives us $s=S(x)$, or $x(s)=S^{-1}(s)$, where
$S^{-1}$ is the inverse function of $S$.
If $f=\exp$, $S(x)=1-\exp(-x)$ and we confirm that
$x(s)=-\log(1-s)$.

We can use this result to 
calculate the derivatives of a function $x(s)$,
given as the inverse of a function $S(x)$.
To do this, we define $f(x)=1/S'(x)$ and, using
Eq.(\ref{Butcherflow}), we obtain
\begin{eqnarray}
x^{(n)}(0) &=&\sum_{|t|=n} \alpha(t) \delta_t,
\end{eqnarray}
where $\delta_t$ is calculated from $f(s)$ using 
Eq.(\ref{defdeltatsimple}) in section \ref{sectiondeltat}.

This method can also be calculated to find the function $f$
satisfying given values for 
\begin{eqnarray*}
a_n=\sum_{|t|=n} \alpha(t) \delta_t,
\end{eqnarray*}
where $\delta_t$ is calculated from $f$. For instance,
if we want 
\begin{eqnarray*}
\sum_{|t|=n} \alpha(t) \delta_t=n!,
\end{eqnarray*}
we must take $f(s)=1+s^2$.

\subsubsection{Other sums over trees \label{sectionothersum}}
We give now further examples of sums over trees,
that will be used in the sequel.
For instance, assume that we need to compute
\begin{eqnarray*}
S=\sum_{|t|=n} \frac{\alpha(t)}{t!}.
\end{eqnarray*}
This term comes in the Butcher series with 
$\phi(t)=1/(t!)^2$. Since this $\phi(t)$ is the square
of the previous one, the corresponding Runge-Kutta
method can be realized as
 the tensor product of two ``simple integral methods''
(see section \ref{Butcherexample}).
In other words
\begin{eqnarray*}
a(u,u',v,v')&=&a(u,v) a(u',v')=1_{[0,u]}(v)1_{[0,u']}(v')
\quad b(u,u')=b(u)b(u')=1.
\end{eqnarray*}
and the Runge-Kutta method is now
\begin{eqnarray*}
X_{uu'}(s)&=&x_0+(s-s_0)\int_0^u dv \int_0^{u'} dv'f[X_{vv'}(s)] \\
x(s)&=&x_0+(s-s_0)\int_0^1du \int_0^1du' f[X_{uu'}(s)].
\end{eqnarray*}
The corresponding homomorphism $\phi(t)$ is given by
\begin{eqnarray*}
\phi_{uu'}(\tun)&=&1\\
\phi_{uu'}(B^+(t_1\cdots t_k))&=&\int_0^u du_1 \int_0^{u'} du'_1 
       \phi_{u_1u'_1}(t_1)
       \dots\int_0^u du_k \int_0^{u'} du'_k \phi_{u_ku'_k}(t_k)\\
\phi(t)&=&\int_0^1 du \int_0^1 du'\phi_{uu'}(t).
\end{eqnarray*}
The solutions of these equations are
\begin{eqnarray*}
\phi_{uu'}(t)&=&\frac{|t|^2(uu')^{|t|-1}}{(t!)^2}\\
\phi(t)&=&\frac{1}{(t!)^2},
\end{eqnarray*}
so that, from Eq.(\ref{Bseries})
\begin{eqnarray*}
X_{uu'}(s)&=&x_0+\sum_t \frac{(s-s_0)^{|t|}}{|t|!}
  \frac{\alpha(t)(uu')^{|t|}}{t!} \delta_t(s_0). 
\end{eqnarray*}
The conclusion is that $X_{uu'}(s)$ is in fact a function
of $uu'$ and not of $u$ and $u'$.
More precisely, we know from the general formula Eq.(\ref{Bseries})
that the B-series for the solution of 
$x(s)=x_0+(s-s_0)\int_0^1du \int_0^1du' f[X_{uu'}(s)]$ is
\begin{eqnarray*}
x(s)&=&x_0+ \sum_t \frac{(s-s_0)^{|t|}}{|t|!}
  \frac{\alpha(t)}{t!} \delta_t(s_0),
\end{eqnarray*}
so that $X_{uu'}(s)=x(s_0+(s-s_0)uu')$.
If we use the successive changes of variables $w=uu'$,
$v'=s_0+(s-s_0)w$ and $v=s_0+(s-s_0)u$ we find
\begin{eqnarray*}
x(s)&=&x_0+(s-s_0)\int_0^1du \int_0^1du' f[x(s_0+(s-s_0)uu')]\\
&=&x_0+(s-s_0)\int_0^1\frac{du}{u} \int_0^u dw f[x(s_0+(s-s_0)w)]\\
&=&x_0+\int_0^1\frac{du}{u} \int_{s_0}^{s_0+(s-s_0)u} 
  dv' f[x(v')]\\
&=&x_0+\int_{s_0}^{s}\frac{dv}{v-s_0} \int_{s_0}^{v} 
  dv' f[x(v')].
\end{eqnarray*}
With the initial values $x_0=s_0=0$ this gives us
\begin{eqnarray}
x(s) &=&\int_{0}^{s}\frac{dv}{v} \int_{0}^{v} dv' f[x(v')], 
\label{doublefact}
\end{eqnarray}
or
$sx''+x'=f(x)$ with $x(0)=0$ and $x'(0)=f(0)$.
If we take again
$f(x)=\exp(x)$ we find
$sx''+x'=\exp(x)$ with $x(0)=0$ and $x'(0)=1$, so that
\begin{eqnarray*}
x(s)=-2\log(1-s/2)=\sum_{n=1}^\infty \frac{s^n}{n2^{n-1}}.
\end{eqnarray*}
Comparing this with the B-series
\begin{eqnarray*}
x(s)&=&
\sum_t \frac{s^{|t|}}{|t|!}
  \frac{\alpha(t)}{t!}
\end{eqnarray*}
we obtain
\begin{eqnarray*}
S=\sum_{|t|=n} \frac{\alpha(t)}{t!}=\frac{(n-1)!}{2^{n-1}},
\end{eqnarray*}
which is the result found by Kreimer in \cite{Kreimer}
using combinatorial arguments.

As a final example, we can consider the Runge-Kutta method
$a(u,v)=1$, $b(u)=1$ which gives $\phi(t)=1$ for all trees $t$.
The equation for $x(s)$ is now a fixed point problem
$x(s)=s\exp(x(s))$, whose well-known solution is
\begin{eqnarray*}
x(s)&=&
\sum_n \frac{s^n}{n!} n^{n-1},
\end{eqnarray*}
so that
\begin{eqnarray*}
\sum_{|t|=n} \alpha(t) t!=n^{n-1}.
\end{eqnarray*}

These examples show that B-series can be used as
generating series for sums over trees.

\subsection{The antipode \label{sectionantipode}}
The Hopf algebra structure of the ART entails an antipode $S$.
If $\phi(t)$ is an homomorphism, the action of the antipode
on $\phi$ can be written as
$S(\phi)(t)=\phi(S(t))$. 
If the Runge-Kutta method
for $\phi$ is $A_u$, $B$, then the Runge-Kutta method for
$\phi^S=S(\phi)$ is $A^S_u=A_u-B$, $B^S=-B$.
It is useful to see it working on simple cases:
\begin{eqnarray*}
\phi^S_u(\tun)&=&1=\phi_u(\tun)\\
\phi^S(\tun)&=&B^S(\phi^S_u(\tun))=-B(\phi_u(\tun))=-\phi(\tun)\\
\phi^S_u(\tdeux)&=&A^S_u(\phi^S_v(\tun))=A_u(\phi_v(\tun))-B(\phi_v(\tun))=
\phi_u(\tdeux)-\phi(\tun)\\
\phi^S(\tdeux)&=&-B(\phi^S_u(\tdeux))=-\phi(\tdeux)+B(1)\phi(\tun)=
  -\phi(\tdeux)+\phi(\tun)\phi(\tun)
\end{eqnarray*}

\subsection{The convolution \label{sectionconvolution}}
The convolution of $\phi$ and $\phi'$ is defined as
$\phi''(t)=(\phi\star\phi')(t)=m[(\phi\otimes\phi')\Delta(t)]$.

Let $A_u$, $B$ and $A'_u$, $B'$ be the Runge-Kutta methods of,
respectively, $\phi(t)$ and $\phi'(t)$. To be specific, we
consider that $u$ varies from 0 to 1. Then 
the Runge-Kutta method
for $\phi''$ is  $A''_u$, $B''$, where $u$ varies from
0 to 2 and
\begin{eqnarray*}
A''_u(X_v)&=&A_u(X_v)\quad\mathrm{if}\quad 0\le u\le 1
  \quad\mathrm{and}\quad 0\le v\le 1\\
A''_u(X_v)&=&0\quad\mathrm{if}\quad 0\le u\le 1
  \quad\mathrm{and}\quad 1\le v\le 2\\
A''_u(X_v)&=&B(X_v) \quad\mathrm{if}\quad 1\le u\le 2
  \quad\mathrm{and}\quad 0\le v\le 1\\
A''_u(X_v)&=&A'_{u-1}(X_{v-1}) \quad\mathrm{if}\quad 1\le u\le 2
  \quad\mathrm{and}\quad 1\le v\le 2\\
B''(X_v)&=&B(X_v)\quad\mathrm{if}\quad 0\le v\le 1\\
B''(X_v)&=&B'(X_{v-1})\quad\mathrm{if}\quad 1\le v\le 2.
\end{eqnarray*}

Again, we show the formula in action:
\begin{eqnarray*}
\phi''_u(\tun)&=&1\\
\phi''(\tun)&=&B(1)+B'(1)=\phi(\tun)+\phi'(\tun)\\
\phi''_u(\tdeux)&=&A''_u(\phi''_v(\tun))=A_u(1) 1_{[0,1](u)}
  +(B(1)+A'_{u-1}(1))1_{[1,2](u)}\\
\phi''(\tdeux)&=&B(A_u(1))+B'((B(1)+A'_{u-1}(1)))=
  B(\phi_u(\tun))+B(1)B'(1)+B'(\phi'u(\tun))\\
  &=& \phi(\tdeux)+\phi(\tun)\phi'(\tun)+\phi'(\tdeux).
\end{eqnarray*}

\section{Runge-Kutta methods for renormalization}

In this section, we shall follow closely Kreimer's paper 
\cite{Kreimer} and define, for each operation on homomorphisms,
a corresponding transformation of the Runge-Kutta methods.
Instead of attempting a general theory, we consider a
specific example in detail.

\subsection{Runge-Kutta method for bare quantities}

We consider that a given bare physical quantity can be
calculated as a sum over trees, and that the corresponding
Runge-Kutta method has been found as
a pair of linear operators $A_u$ and $B$. The usual 
combinatorial proof show that the solution of the equations
(we take $s_0=0$)
\begin{eqnarray*}
X_u(s) &=& x_0+s A_u[f(X_v(s))]\\
x(s) &=& x_0+s B[f(X_u(s))],
\end{eqnarray*}
is 
\begin{eqnarray*}
X_u(s)&=&x_0+\sum_t \frac{s^{|t|}}{|t|!}
  \alpha(t) t! A_u[\phi_v(t)] \delta_t\\
x(s)&=&x_0+\sum_t \frac{s^{|t|}}{|t|!}
  \alpha(t) t! \phi(t) \delta_t,
\end{eqnarray*}
where, as usually,
\begin{eqnarray*}
\phi_u(\tun)&=&1\\
\phi_u(B^+(t_1\cdots t_k))&=&A_u[\phi_{u_1}(t_1)]
       \dots A_u[\phi_{u_k}(t_k)]\\
\phi(t)&=&B[\phi_u(t)].
\end{eqnarray*}
Here $x(s)$ is the sum giving the bare quantity of interest.
In the examples developed by Broadhurst and Kreimer \cite{Broadhurst},
the quantity of interest is
\begin{eqnarray*}
x(s)&=&\sum_t \frac{s^{|t|}}{|t|!} B_t,
\end{eqnarray*}
where $B_t$ is obtained recursively from given $B_n$ by 
\begin{eqnarray}
B_\bullet &=& B_1 \nonumber\\
B_t &=& B_{|t|} \delta_{t_1} \delta_{t_2} \cdots \delta_{t_k}
\quad\mathrm{when}\quad t=B^+(t_1,t_2,\cdots,t_k). 
\end{eqnarray}

In the renormalization problems considered by Broadhurst and Kreimer,
the $B_n$ are defined from a function
$L(\delta)$ regular (and equal to 1) at the origin, by
\begin{eqnarray*}
B_n &=&\frac{L(n\epsilon)}{n\epsilon}.
\end{eqnarray*}

A pair of operators giving $\phi(t)=B_t$ can be defined as
\begin{eqnarray*}
A_u(X_v) &=& \frac{1}{\epsilon} \int_0^u L(\epsilon \frac{d}{dv}v)X_v,
\quad B(X_v)=A_1(X_v).
\end{eqnarray*}

The quantity of interest $x(s)$ is then obtained by tensoring
$A_u$ with the ``simple integral method'' to obtain $\phi(t)=B_t/t!$.

The only thing that we need in the following is the action of
$A_u$ on a monomial $v^{n-1}$
\begin{eqnarray}
A_u(v^{n-1}) &=& \frac{1}{\epsilon} \int_0^u L(\epsilon \frac{d}{dv}v)v^{n-1}=
\frac{1}{\epsilon} \int_0^u v^{n-1}dv L(n \epsilon )=B_n u^n. \label{actiondeAu}
\end{eqnarray}

\subsection{$S_R$, the ``renormalized antipode''}
In ref.\cite{Kreimer}, Kreimer defines recursively 
a renormalized antipode\footnote{In Hopf algebra terms
$S_R(\phi)(t)=-R[\phi(t)+m[(S_R\otimes Id)(\phi\otimes \phi)P_2\Delta(t)]$.}
depending on a renormalization scheme $R$.
We take as an example the toy model used by Kreimer, where
$R[\phi]=\langle\phi\rangle$ is the projection of $\phi$ on
the pole part of the Laurent series in $\epsilon$ inside the
bracket.

Following the results of section \ref{sectionantipode}, 
the Runge-Kutta method for $S_R(\phi)$ can be obtained from the
Runge-Kutta method of $\phi$ by 
$A^S_u(X)=A_u(X)-\langle A_1(X)\rangle $, 
$B^S(X)=-\langle A_1(X)\rangle $. 
Working out the first examples using Eq.(\ref{actiondeAu}), we find,
\begin{eqnarray*}
\phi^S_u(\tun)&=&1\\
\phi^S(\tun)&=&\langle A_1(1) \rangle=\langle B_1 \rangle\\
\phi^S_u(\tdeux)&=&A_u(\phi_v(\tun))-\langle A_1(\phi_v(\tun))\rangle=
       A_u(1)-\langle A_1(1)\rangle=B_1 u- \langle B_1 \rangle\\
\phi^S(\tdeux)&=&-\langle A_1(\phi^S_u(\tdeux))\rangle=
   -\langle B_2 B_1 \rangle + \langle\langle B_1 \rangle B_1 \rangle.
\end{eqnarray*}

\subsection{Renormalized quantities}
Finally, the renormalized quantities $x^R(s)$ are obtained from the
convolution of $S_R(\phi)$ with $\phi$.
To obtain the corresponding Runge-Kutta method, we use
the results of section \ref{sectionconvolution}. However,
the domain where $1\le u\le 2$ is not used, and the 
Runge-Kutta method for the renormalized quantity is
$A^R_u(X)=A_u(X)-\langle A_1(X)\rangle $,
$B^R(X)=A_1(X)-\langle A_1(X)\rangle $.
It may seem surprising that such a simple equation
encodes the full combinatorial complexity of renormalization.
It is not even necessary to work examples out, because 
$A^R_u(X)=A^S_u(X)$ so that $\phi^R_u(t)=\phi^S_u(t)$, 
and the only difference comes from the action of $B^R$.

For a real calculation of $x^R(s)$,
we do not need $A^R_u$ and $B^R$ which give us $\phi(t)=\Gamma(t)$,
but the tensor product of this
method with the ``simple integral method'' to obtain 
$\phi(t)=t! \Gamma(t)$.
In detail, the equation for the renormalized quantity 
$x^R(s)$ is
\begin{eqnarray}
X^R_{uu'}(s)&=&\frac{s}{\epsilon}\int_0^u dv \int_0^{u'} dv'
 L(\epsilon \partial_v v) e^{X_{vv'}(s)}-
\langle \frac{s}{\epsilon}\int_0^1 dv \int_0^{u'} dv'
 L(\epsilon \partial_v v) e^{X_{vv'}(s)}\rangle \\
x^R(s)&=&\frac{s}{\epsilon}\int_0^1 dv \int_0^{1} dv'
 L(\epsilon \partial_v v) e^{X_{vv'}(s)}-
\langle \frac{s}{\epsilon}\int_0^1 dv \int_0^{1} dv'
 L(\epsilon \partial_v v) e^{X_{vv'}(s)}\rangle.
\label{Req1}
\end{eqnarray}

For a general renormalization scheme $R$, one replaces 
$\langle A_u(X) \rangle$ by $R[A_u(X)]$. Finally, Chen's
lemma for renormalization schemes \cite{Kreimer} is 
obtained from Hairer and Wanner's theorem Eq.(\ref{Hairer2}).

\section{Renormalization of Kreimer's toy model}

In this section, we use Runge-Kutta methods to renormalized
explicitly Kreimer's toy model for even functions $L(\epsilon)$.
In \cite{Broadhurst}, remarkable properties of the renormalized sum 
of diagrams with ``Connes-Moscovici weights'' were noticed.

\subsection{Equation for the renormalized quantity}
The role of the sum over $u'$ in Eq.(\ref{Req1}) is 
to add a factor $1/t!$, as in
section{\ref{sectionothersum}}. Therefore, the same reasoning
can be used to show that $X^R_{uu'}(s)$ is in fact a function 
of $su'$ and we write $X^R_{uu'}(s)=X^R_{u}(su')$, which defines
the function $X^R_{u}(s)$. The equation for $X^R_{u}(s)$ can be found
from  Eq.(\ref{Req1}) and the relation $X^R_{u}(s)=X^R_{us}(1)$ as

\begin{eqnarray}
X^R_{u}(s)&=&\frac{1}{\epsilon}\int_0^u dv \int_0^{s} ds'
 L(\epsilon \partial_v v) e^{X_{v}(s')}-
\langle \frac{1}{\epsilon}\int_0^1 dv \int_0^{s} ds'
 L(\epsilon \partial_v v) e^{X_{v}(s')}\rangle \\
x^R(s)&=&\frac{1}{\epsilon}\int_0^1 dv \int_0^{s} ds'
 L(\epsilon \partial_v v) e^{X_{v}(s')}-
\langle \frac{1}{\epsilon}\int_0^1 dv \int_0^{s} ds'
 L(\epsilon \partial_v v) e^{X_{v}(s')}\rangle.
\label{Req2}
\end{eqnarray}
To solve this equation, we expand $X^R_{u}(s)$ in a power series
over $u$:
\begin{eqnarray*} 
X^R_{u}(s)&=&\sum_{n=0}^{\infty} a_n(s)u^n.
\end{eqnarray*} 
A standard identity gives us
\begin{eqnarray*} 
\exp(X^R_{u}(s))&=&\sum_{n=0}^{\infty} \lambda_n(a)u^n,\quad\mathrm{where}\\
\lambda_n(a)&=&\sum_{|\alpha|=n}\frac{a_1^{\alpha_1}\cdots a_n^{\alpha_n}}
              {{\alpha_1}!\cdots{\alpha_n}!}\quad\mathrm{with}\quad
 |\alpha|=a_1+2\alpha_2+\dots+n\alpha_n.
\end{eqnarray*} 
$\lambda_n(a)$ depends on $s$ through its arguments $a_i(s)$.
The sets of $\alpha_i$ for a given $n$ can be obtained from the partitions
of $n$: $(\mu_1,\dots,\mu_n)$, where 
$\mu_1 \ge \cdots \ge \mu_n$ by $\alpha_n=\mu_n$,
$\alpha_i=\mu_i-\mu_{i+1}$ for $i<n$.

The first $\lambda_n(a)$ are
\begin{eqnarray*} 
\lambda_0(a)&=&1\quad
\lambda_1(a)=a_1\quad
\lambda_2(a)=a_2+\frac{a_1^2}{2}\quad
\lambda_3(a)=a_3+a_1a_2+\frac{a_1^3}{6}.
\end{eqnarray*} 

\subsection{Solution of the equation}
Introducing the series expansions for $X^R_{u}(s)$ and
$\exp(X^R_{u}(s))$ into Eq.(\ref{Req2}) we obtain
\begin{eqnarray*}
\sum_{n=0}^{\infty} a_n(s)u^n&=&\sum_{n=0}^{\infty} B_{n+1}
  \int_0^s e^{a_0(s')} \lambda_n(a) ds' u^{n+1}-
\langle\sum_{n=0}^{\infty} B_{n+1}
  \int_0^s e^{a_0(s')} \lambda_n(a) ds' \rangle
\end{eqnarray*}
or
\begin{eqnarray}
a_0(s)&=&-\langle\sum_{n=0}^{\infty} B_{n+1}
  \int_0^s e^{a_0(s')} \lambda_n(a) ds' \rangle\nonumber\\
a_n(s)&=&B_n \int_0^s e^{a_0(s')} \lambda_{n-1}(a) ds'\quad\mathrm{for}\quad n>0.
\label{eqa0an}
\end{eqnarray}

To solve this equation, we need to go back to the equation for the
bare quantity
\begin{eqnarray}
X^0_{u}(s)&=&\frac{1}{\epsilon}\int_0^u dv \int_0^{s} ds'
 L(\epsilon \partial_v v) e^{X^0_{v}(s')}.
\label{eqbare}
\end{eqnarray}
Again $X^0_{u}(s)$ is a function of $su$, we define 
$X^0(s)=X^0_{s}(1)$ which satisfies
\begin{eqnarray*}
X^0(s)&=&\frac{1}{\epsilon}\int_0^s \frac{du}{u} \int_0^{u} dv
 L(\epsilon \partial_v v) e^{X^0(v)}.
\end{eqnarray*}

The solution of this equation is given by the B-series
\begin{eqnarray}
X^0(s)&=&\sum_n {\bar{\alpha}}_n s^n\quad\mathrm{with}\quad
{\bar{\alpha}}_n = \sum_{|t|=n} \frac{\alpha(t)B_t}{|t|!}.
\label{bareeq}
\end{eqnarray}
On the other hand, we can also expand $e^{X^0(v)}$ using
the functions $\lambda_n(\bar{a})$. Identifying both sides of
Eq.(\ref{bareeq}), we obtain the relation
\begin{eqnarray}
{\bar{a}}_n&=&\frac{B_n}{n}\lambda_{n-1}(\bar{a}).\label{abar}
\end{eqnarray}

With this identity, we can now prove that, for the renormalized quantities,
\begin{eqnarray}
a_n(s)&=&(g(s))^n{\bar{a}}_n,\quad\mathrm{where}\quad g(s)=\int_0^s\exp(a_0(s'))ds'.
\label{aid}
\end{eqnarray}
Since $\lambda_0(a)=1$ and ${\bar{a}}_n=B_1$, this equation is true for $n=1$, from
Eq.(\ref{eqa0an}). If Eq.(\ref{aid}) is true up to $n-1$, then 
$\lambda_{n-1}(a)=(g(s))^{n-1}\lambda_{n-1}(\bar{a})$ and 
the derivative of Eq.(\ref{eqa0an}) gives us 
\begin{eqnarray*}
a'_n(s)&=&B_n e^{a_0(s)} \lambda_{n-1}(a)=
B_n g'(s) (g(s))^{n-1} \lambda_{n-1}(\bar{a})=n(g(s))^{n-1} \bar{a}_n,
\end{eqnarray*}
by Eq.(\ref{abar}). Integrating this equation with the condition $a_n(s)=0$
gives Eq.(\ref{aid}) at level $n$.

By this we have proved that the flow for the renormalized quantity is
a reparametrization of the flow for the bare quantity:
$X^R_u(s)=a_0(s)+X^0(u g(s))$ and
$X^R(s)=a_0(s)+X^0(g(s))$.

To determine $a_0(s)$ we proceed step by step. In Eq.(\ref{bareeq}) we
expand $L(\epsilon \partial_v v)$ over $\epsilon$.
The first term is just 1, and we obtain Eq.(\ref{doublefact}) with
the solution $x(s)=-2\log(1-s/(2\epsilon))$.
For the renormalized quantity, the most singular term
becomes $X^0(g(s))=-2\log(1-g(s)/(2\epsilon))$. Since $X^R(s)$ is regular,
this singular term must be compensated by a corresponding term in
$a_0(s)$. By equating the most singular terms we obtain
$a_0(s)=-2\log(1-g(s)/(2\epsilon))$. We know from Eq.(\ref{aid}) that
$a_0(s)=\log(g'(s))$, and we obtain the most singular terms as
the solution of $g'(s)=1/(1-g(s)/(2\epsilon))^2$, which is:
\begin{eqnarray*}
g(s)&=&\frac{s}{1+\frac{s}{2\epsilon}}\\
a_0(s)&=& -2\log(1+\frac{s}{2\epsilon}).
\end{eqnarray*}
By expanding $a_0(s)$ as a series in $s$, we obtain the most singular
term observed in \cite{Broadhurst} and proved in \cite{Kreimer}.
One notices that the singularity of the non-pertubative term
$a_0(s)$ is logarithmic, and much smoother than the singularities coming
from the expansion over $s$ (i.e. the perturbative expression).

\subsection{Differential equation for the finite part}
In general, one should proceed now with the next singular term.
To obtain it we denote $Y(s)=X^0(g(s))$, this change of variable gives 
the equation for $Y(s)$:
\begin{eqnarray*}
Y(s)&=&\frac{1}{\epsilon}\int_0^s \frac{g'(u)du}{g(u)} \int_0^{u} dv
 g'(v) L(\epsilon+\epsilon \frac{g(v)}{g'(v)}
 \partial_v ) e^{Y(v)}.
\end{eqnarray*}

Now we can write $Y(s)=X^R(s)-a_0(s)$, and notice that the term
$-a_0(s)$ on the left-hand side is compensated by a term on the
right-hand side where $L=1$ and $\exp(X^R(s))=1$. 
We obtain the equation for $X^R(s)$:
\begin{eqnarray*}
X^R(s)&=&\frac{1}{\epsilon}\int_0^s \frac{du}{u(1+\frac{u}{2\epsilon})} 
\int_0^{u} dv\left[
\frac{1}{(1+\frac{v}{2\epsilon})^2}
 L(\epsilon 
 \partial_v v+\frac{v^2}{2} 
 \partial_v){(1+\frac{v}{2\epsilon})}^2 e^{X^R(v)}
 -1\right].
\end{eqnarray*}
The nice aspect of the previous equation is that it seems to
have a limit as $\epsilon$ goes to zero. In fact, it has a limit
when $L$ is even, as we shall show now.

Writing $\bar{X}(s)=\lim_{\epsilon\rightarrow 0}X^R(s)$, and
taking the limit $\epsilon\rightarrow 0$ in the previous equation,
we obtain
\begin{eqnarray*}
\bar{X}(s)&=&2\int_0^s \frac{du}{u^2} 
\int_0^{u} dv\left[
\frac{1}{v^2}
 L(\frac{v^2}{2} \partial_v )v^2 e^{\bar{X}(v)}
 -1\right],
\end{eqnarray*}
or, in differential form:
\begin{eqnarray*}
\frac{1}{2}(s^2\bar{X}'(s))'&=&\frac{1}{s^2}
 L(\frac{s^2}{2}\frac{d}{ds})s^2 e^{\bar{X}(s)}
 -1.
\end{eqnarray*}

If $\bar{X}(s)$ and $ L(\delta)$ are expanded as
\begin{eqnarray}
\bar{X}(s)&=& \sum_{n=1}^\infty b_n s^n\quad\mathrm{and}\quad
 L(\delta)= 1+\sum_{n=1}^\infty L_n \delta^n,\quad\mathrm{so that} \nonumber\\
 L(\frac{s^2}{2}\frac{d}{ds})&=& 1+\sum_{n=1}^\infty L_n (\frac{s^2}{2}\frac{d}{ds})^n,
\label{renoreq}
\end{eqnarray}
we obtain the following relation for the term in $s$:
$b_1s=(b_1+L_1/2)s$. If $L_1$ is not zero, we obtain a contradition and
must proceed with the withdrawal of divergences. For simplicity,
we shall assume that $L_1=0$. Then $b_1$ becomes a free parameter
of $\bar{X}(s)$. All terms $b_n$ with $n>1$ can now be
determined from $b_1$ and $L_n$ ($n>1$). All terms are regular.

In \cite{Broadhurst}, the function $L(\delta)$ was taken
even. Then $L_1=0$, and their results correspond to $b_1=0$.
Broadhurst and Kreimer have also used a function
$L(\epsilon,\delta)$. The present treatment can be
applied to this more general situation, with the 
only change that 
\begin{eqnarray*}
L_n=n!\lim_{\epsilon\rightarrow 0}\lim_{\delta\rightarrow 0} 
\frac{d^n}{d\delta^n} L(\epsilon,\delta).
\end{eqnarray*}

Clearly, Eq.(\ref{renoreq}) is much faster to solve than computing the sum
over trees. For instance, the expansion could be calculated up
to 20 loops (i.e. $b_{20}$) within a few seconds with a computer.

\subsection{Alternative point of view}
There is an alternative way to solve Eq.(\ref{eqbare}) for the
bare quantity. We define a function $f(s)$ from $L(\delta)$ by
\begin{eqnarray*}
f(s)&=& \sum_{n=0}^\infty \frac{L(n\epsilon+\epsilon)}{n!} s^n=
L(\epsilon \frac{d}{ds}s) e^s.
\end{eqnarray*}
A relation between $f(s)$ and $L(\delta)$ can also be established
through the Mellin transforms of $f$ and $L$ as
$M(f)(z)=M(L)(\epsilon-\epsilon z)\Gamma(z)$.

With $f(s)$ we can write the equation for the bare quantity as
\begin{eqnarray}
X^0(s)&=&\frac{1}{\epsilon}\int_0^s \frac{du}{u} \int_0^{u} dv
f(X^0(v)). \label{eqavecf}
\end{eqnarray}

Alternatively, one can go from $f$ to $L$ and
consider the results of the toy model as a method to 
renormalize equations of the type (\ref{eqavecf}).

\section{n-dimensional problems}
For applications to classical field theory, we need to develop Runge-Kutta
methods for the n-dimensional analogue of the flow equation:
non-linear partial differential equations.
The purpose of the present section is to indicate how B-series
can be used for this case\footnote{Kreimer was independently aware
of the possibility to use B-series for non-linear partial differential
equations.}.
The method apply to equations of the form
$L\psi(\mathbf{r})= F[\psi(\mathbf{r})]$,
where $L$ is a differential operator (e.g. 
the nonlinear Schr\"odinger equation $\Delta\psi=\psi^3$).

\subsection{Formulation}
We need two starting elements: a function $\psi_0(\mathbf{r})$
which is the solution of $L\psi_0(\mathbf{r})=0$, 
and a Green function $G(\mathbf{r},\mathbf{r}')$,
that is a solution of the equation
$L_rG(\mathbf{r},\mathbf{r}')=\delta(\mathbf{r}-\mathbf{r}')$,
with given boundary conditions. The function $\psi_0(\mathbf{r})$ 
will play the role of an initial value, and the Green function
will decide in which ``direction'' you move from the initial
value. It will also state, in some sense, the boundary conditions
of the solution $\psi(\mathbf{r})$.

Using these two functions, the differential equation
$L\psi(\mathbf{r})=F[\psi(\mathbf{r})]$ is transformed
into 
$\psi(\mathbf{r})=\psi_0(\mathbf{r})+\int d\mathbf{r}'
G(\mathbf{r},\mathbf{r}')F[\psi(\mathbf{r}')]$. The action
of $L$ enables us to go from the second to the first equation.

The combinatorics is the same as for the standard
Runge-Kutta method, and the result is
\begin{eqnarray}
\psi(\mathbf{r})&=&\psi_0(\mathbf{r})+\sum_t \frac{\alpha(t) t!}{|t|!}
  \int d\mathbf{r}' G(\mathbf{r},\mathbf{r}')
  \phi_{r'}(t),\label{Bseries_n}
\end{eqnarray}
where $\phi_{r}(t)$ is defined recursively by
\begin{eqnarray}
\phi_{r}(\tun)&=&F[\psi_0(\mathbf{r})]\nonumber\\
\phi_{r}(B^+(t_1\cdots t_k))&=&F^{(k)}[\psi_0(\mathbf{r})]
 \int dr_1 G(\mathbf{r},\mathbf{r}_1)\phi_{r_1}(t_1)
       \dots \int dr_k G(\mathbf{r},\mathbf{r}_k)\phi_{r_k}(t_k). \label{Bphi_n}
\end{eqnarray}

If $\psi$ is a vector field, the solution is the same,
and equations (\ref{Bphi_n}) get indices:
\begin{eqnarray*}
\phi^i_{r}(\tun)&=&f^i[\psi_0(\mathbf{r})]\\
\phi^i_{r}(B^+(t_1\cdots t_k))&=&F^i_{j_1\dots j_k}[\psi_0(\mathbf{r})]
 \int dr_1 G^{j_1}_{j'_1}(\mathbf{r},\mathbf{r}_1)\phi^{j'_1}_{r_1}(t_1)
       \dots \int dr_k G^{j_k}_{j'_k}(\mathbf{r},\mathbf{r}_k)\phi^{j'_k}_{r_k}(t_k),
\end{eqnarray*}
where $G^i_j(\mathbf{r},\mathbf{r}')$ is a component of the matrix Green function.

In the previous sections, the series (\ref{Bseries})
was written as a function of $\phi(t)$ 
(describing the effect of the Runge-Kutta method ($a$,$b$)) 
and $\delta_t$ (describing the effect of the
function $F[x]$). In the present case, this separation is no longer possible,
and $\phi(t)$ combines both pieces of information.

\subsection{Examples}
In this section, equation (\ref{Bseries_n}) is applied to the 
one-dimensional problem and to the Schr\"odinger equation.

\subsubsection{The one-dimensional case}
It is instructive to observe how the one-dimensional case is
obtained from Eq.(\ref{Bseries_n}). The differential operator
is $L=d/ds$, so the initial function $\psi_0(s)$ must satisfy
$d/ds\psi_0(s)=0$: $\psi_0(s)$ is a constant that we write $x_0$.
For the Green function $G(s,s')$, we have the equation
$LG(s,s')=\delta(s-s')$, so $G(s,s')=\theta(s-s') +C(s')$,
where $\theta(s)$ is the step function and $C(s')$ a
function of $s'$. To determine $C(s')$,
we note that, in the ``simple integral method'', 
there is an integral from $s_0$ to $s$.
From the Green function 
$G(s,s')=\theta(s-s') -\theta(s_0-s')$, we obtain 
\begin{eqnarray*}
\int_{-\infty}^{\infty} G(s,s') f(s') ds'&=& \int_{s_0}^{s} f(s') ds'
\end{eqnarray*}
which is the required expression. 

Now, the role of $\psi_0$ and the Green function is clear for the
one-dimensional case: $\psi_0$ gives the initial value $x_0$ and 
$G$ specifies (among other things) the starting point $s_0$.
To complete the derivation of the one-dimensional case, we note
that $\psi_0(s)=x_0$ does not depend on $s$, so the
terms $F^{(k)}[\psi_0(s)]=F^{(k)}[x_0]$ are independent of $s$
and can be grouped together to build $\delta_t$ as in (\ref{defdeltat}). 
On the other hand, the integration over Green functions build up
$(s-s_0)^{|t|}/t!$ and we obtain Eq.(\ref{Butcherflow}).

\subsubsection{The Schr\"odinger equation I}
If we write the Schr\"odinger equation as 
$(E+\Delta)\psi(\mathbf{r})=V(\mathbf{r})\psi(\mathbf{r})$,
we can apply Eq.(\ref{Bseries_n}) with 
$F[\psi]=V(\mathbf{r})\psi$. We take for $\phi_0(\mathbf{r})$
a solution of $(E+\Delta)\psi_0(\mathbf{r})=0$ and for
$G(\mathbf{r},\mathbf{r}')$ the scattering Green function
(e.g. $G(\mathbf{r}-\mathbf{r}')=-e^{i\sqrt{E}|\mathbf{r}-\mathbf{r}'|}/(4\pi 
|\mathbf{r},\mathbf{r}'|)$ in three dimensions).

The calculation of $\phi(t)$ is straightforward because, in a
such a linear problem, $F^{(k)}=0$ for $k>1$. Hence, the only
rooted trees that survive are those with one branch.
For these trees $\alpha(t)=1$ and $t!=|t|!$ and we obtain
\begin{eqnarray*}
\psi(\mathbf{r})=\psi_0(\mathbf{r})+\int d\mathbf{r}_1
 G(\mathbf{r},\mathbf{r}_1) V(\mathbf{r}_1)\psi_0(\mathbf{r}_1)+
 \int d\mathbf{r}_1 d\mathbf{r}_2
 G(\mathbf{r},\mathbf{r}_1) V(\mathbf{r}_1)
 G(\mathbf{r}_1,\mathbf{r}_2) V(\mathbf{r}_2)\psi_0(\mathbf{r}_2)+\cdots
\end{eqnarray*}
where we recognize the Born expansion of the Lippmann-Schwinger
equation.

\subsubsection{The Schr\"odinger equation II}
We can also treat the Schr\"odinger equation in an
alternative way as the system of equations:
\begin{eqnarray*}
(E+\Delta)\psi(\mathbf{r})&=& V(\rho)\psi(\mathbf{r}) \\
\frac{\partial \rho_i}{\partial r_j} &=& \delta_{ij}.
\end{eqnarray*}
This is a matrix differential equation. We give index 0 to the
first line, and index $i$ (running from 1 to the dimension of space)
to the other lines, called the space lines.
The purpose of the space lines is just to ensure that 
$\rho=\mathbf{r}$. This is a standard trick to take the
$\mathbf{r}$ dependence of $V$ into account in the expansion
(see e.g. \cite{Hairer} p.143).
As initial value we take $\psi_0(\mathbf{r})$ and $\rho_0=0$,
the matrix Green function is diagonal and it is equal to the scattering 
wave function for  line 0 and to 
$\theta(r_i-r'_i)-\theta(-r'_i)$ for line $i$.

For $\phi_r(\tun)$, the zero-th component is $V(0)\psi_0(\mathbf{r})$
and the space components are 1, for all the other trees, the
space components are 0 and  the zero-th component of the simplest 
tree is
\begin{eqnarray*}
\phi_r(\tdeux)&=&V(0)^2 \int d\mathbf{r}'
 G(\mathbf{r},\mathbf{r}') \psi_0(\mathbf{r}')+
 \sum_i r_i\partial_i V(0) \psi_0(\mathbf{r})\\
\phi_r(\ttroisun)&=&V(0)^3 \int d\mathbf{r}_1 G(\mathbf{r},\mathbf{r}_1) +
 V(0) \int d\mathbf{r}' G(\mathbf{r},\mathbf{r}') 
 \sum_i r'_i\partial_i V(0) \psi_0(\mathbf{r}')\\
\phi_r(\ttroisdeux)&=&2V(0)\sum_i r_i\partial_i V(0)
 \int d\mathbf{r}'
 G(\mathbf{r},\mathbf{r}') \psi_0(\mathbf{r}')+
\sum_{ij} r_i r_j\partial_i\partial_j V(0).
\end{eqnarray*}
The expressions become more and more complex, but their derivation
is made systematic by the recurrence relation.

\section{Conclusion}
Butcher's approach to Runge-Kutta methods was applied to some
simple renormalization problems. Since Cayley, it is clear 
that the ART is ideally suited to treat differentials. This
was confirmed here by presenting a B-series solution of
a class of non-linear partial differential equations.

The recursive nature of B-series make them computationally
efficient: $\phi_u(t)$ can be obtained by a simple operation
from the $\phi_u(t')$ of smaller order $t'$. This is why
B-series can be automated and implemented in a computer.

Butcher's approach has still much to offer.
In the numerical analysis literature, B-series have
been generalized to treat flow equations on
Lie groups. The main change \cite{Munthe98}
is to replace the algebra of
rooted trees by the algebra of planar trees (also called
ordered trees \cite{Owren}).
The elementary differentials get then a ``quantized 
calculus'' flavor, especially in the definition given
Munthe-Kaas \cite{Munthe95} in terms of commutators with
the vector field $F=f^i\partial_i$ (see also Ginocchio).
Using this generalized ART, extended work has been 
carried  out recently for the numerical solution of differential
equations on Lie groups (see Ref.\cite{Munthe98,Owren} and
the web site {\tt http://www.math.ntnu.no/num/synode}).

B-series have been generalized in other directions, e.g. stochastic differential
equations \cite{Komori} and differential equations
of the type $dy/ds=f(y,z)$, $g(y,z)=0$, which are called
differential algebraic equations \cite{HairerII}.

It is our hope that Butcher's approach can be applied to
quantum field theory.

\section{Acknowledgements}
It is my great pleasure to thank Dirk Kreimer and Alain
Connes for interest, encouragement and discussions.

\section{Appendix}

For further reference, the action of the coproduct and the
antipode on the first few trees are given here.

\subsection{Coproduct}

\begin{eqnarray*}
\Delta 1 &=& 1 \otimes 1 \\
\Delta \tun
&=& \tun\otimes 1 + 1 \otimes
    \tun
\\
\Delta \tdeux
&=& \tdeux\otimes 1 + 1 \otimes
    \tdeux+
    \tun\otimes
    \tun
\\
\Delta 
\ttroisun
&=& \ttroisun\otimes 1 + 1 \otimes
    \ttroisun +
    \tdeux\otimes
    \tun+
    \tun\otimes
    \tdeux
\\
\Delta 
\ttroisdeux
&=& \ttroisdeux\otimes 1 + 1 \otimes
    \ttroisdeux +
    \tun\tun\otimes
    \tun+2
    \tun\otimes
    \tdeux
\\
\Delta 
\tquatreun
&=& \tquatreun\otimes 1 + 1 \otimes
    \tquatreun +
    \tun\otimes
    \ttroisun+
    \ttroisun\otimes
    \tun+
    \tdeux\otimes
    \tdeux
\\
\Delta 
\tquatredeux
&=& \tquatredeux\otimes 1 + 1 \otimes
    \tquatredeux + 2
    \tun\otimes
    \ttroisun+
    \ttroisdeux\otimes
    \tun+
    \tun\tun\otimes
    \tdeux
\\
\Delta 
\tquatretrois
&=& \tquatretrois\otimes 1 + 1 \otimes
    \tquatretrois + 
    \tun\otimes
    \ttroisun+
    \tun\otimes
    \ttroisdeux+
    \tdeux\otimes
    \tdeux+
    \tun\tun\otimes
    \tdeux+
    \tun\tdeux\otimes\tun
\\
\Delta 
\tquatrequatre
&=& \tquatrequatre\otimes 1 + 1 \otimes
    \tquatrequatre + 
    3\tun\otimes
    \ttroisdeux+
    3\tun\tun\otimes
    \tdeux+
    \tun\tun\tun\otimes
    \tun
\end{eqnarray*}

\subsection{Antipode}

\begin{eqnarray*}
S(1) &=& 1 \\
S(\tun)
&=& -\tun
\\
S\left(\tdeux\right)
&=& -\tdeux+ \tun \tun
\\
S\left( \ttroisun\right)
&=& -\ttroisun +2 \tun\tdeux -\tun\tun\tun
\\
S\left( \ttroisdeux\right)
&=& -\ttroisdeux +2 \tun\tdeux -\tun\tun\tun
\\
S\left( \tquatreun\right)
&=& -\tquatreun+2 \tun \ttroisun+
    \tdeux \tdeux-3\tun\tun\tdeux+
    \tun\tun\tun\tun
\\
S\left( \tquatredeux\right)
&=& -\tquatredeux+2 \tun \ttroisun+
    \tun \ttroisdeux-3\tun\tun\tdeux+
    \tun\tun\tun\tun
\\
S\left( \tquatretrois\right)
&=& -\,\tquatretrois+\tun \ttroisun+
    \tun \ttroisdeux+
    \tdeux \tdeux-3\tun\tun\tdeux+
    \tun\tun\tun\tun
\\
S\left( \tquatrequatre\right)
&=& -\tquatrequatre+
    3\tun \ttroisdeux
    -3\tun\tun\tdeux+
    \tun\tun\tun\tun
\end{eqnarray*}

\end{fmffile}

\end{document}